\PassOptionsToPackage{english}{babel}
\documentclass[%
 amsmath,amssymb,
 aps,
prb,
reprint,
]{revtex4-2}

\usepackage{graphicx}% Include figure files
\usepackage{dcolumn}% Align table columns on decimal point
\usepackage{bm}% bold math
\usepackage[english]{babel}
\usepackage{braket}
\usepackage{array}
\usepackage[colorinlistoftodos,prependcaption,textsize=tiny]{todonotes} %option "disable" turns todo off 
\newcommand{\note}[1]{\textcolor{black}{#1}} %black!50!green
\usepackage[normalem]{ulem}
\newcommand{\dotkp}{\boldsymbol{k}\cdot\boldsymbol{p}}
\usepackage{url}
\usepackage{makecell}
\usepackage{lipsum}
\usepackage{amsmath,mathtools}

\begin{document}
\selectlanguage{English}

\preprint{APS/123-QED}

\title{The Emission Directionality of Electronic Intraband Transitions in Stacked Quantum Dots}% Force line breaks with \\

\author{Alexander~Mittelst{\"a}dt}
 \email{alexander.mittelstaedt@physik.tu-berlin.de}
\author{Ludwig~A.~Th.~Greif}%
\author{Andrei~Schliwa}%
\affiliation{%
 Institut f{\"u}r Festk{\"o}rperphysik, Technische Universit{\"a}t Berlin, \\
 Hardenbergstr. 36, 10623 Berlin, Germany
}%

\date{\today}% It is always \today, today,

\begin{abstract}
%
%\lipsum[1]
% 600 Zeichen max
%
We investigate the emission directionality of electronic intraband (intersubband) transitions in stacked coupled quantum dots.
Using a well-established eight-band $\dotkp$ method, we demonstrate that the minor contributions from the valence band mixing into the conduction band govern the polarization and emission directionality of electronic $p$-to-$s$-type intraband transitions.
Despite that the contribution from the central-cell part to the momentum matrix element is dominant, we find, the contribution from the matrix elements among the envelope functions cannot be neglected.
With the help of an artificial cuboidal quantum dot, we show that the vertical emission from intraband transitions can be tuned via the emitter's vertical aspect ratio.
Subsequently, we show that these results can be transferred to more realistic geometries of quantum dots and quantum dot stacks and demonstrate that the vertical emission can be 
enhanced from $23$\,\% to $46$\,\% by increasing the emitter's vertical aspect ratio to the isotropic case with a vertical aspect ratio of 1.0.
Therefore, a stacking of a few quantum dotss ($\sim$4 for the investigated structures) is already sufficient to improve the vertical radiation significantly. 
Additionally, we discuss the impact of the number of stacked QDs as well as the effect of the interdot coupling strength on the radiation properties. 
\end{abstract}

%\keywords{Suggested keywords}%Use showkeys class option if keywordmixing into the conduction band 
\maketitle

%\tableofcontents
%
%
\section{Introduction}
\begin{figure}[tb]
 	\centering
    \includegraphics[]{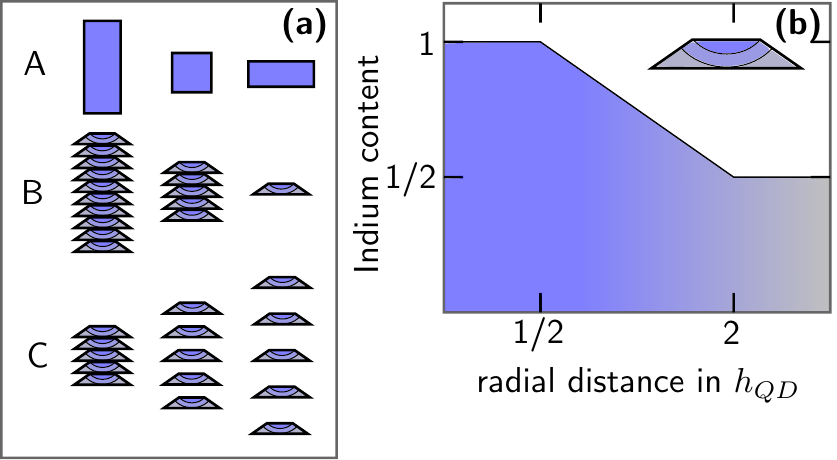}\\
	\caption{(a) The QD structures investigated. Series A: nine artificial cubodial QDs with $AR$ between 1.6 and 0.4. Series B: stacks of ten identical QDs to a single QD with barrier width of $b=0$\,monolayers (MLs). Series C: stacks of nine identical QDs with barriers varying from $b=0$\,MLs to $b=16$\,MLs.
	(b) Scheme of the modeled QD geometry with a gradual composition. The Indium $x$ content within an In$_{x}$Ga$_{1-x}$As QD as a function of the distance to the apex center in units of the QD's height $h_{\textrm{QD}}$.
	}
	\label{fig:QD_model}%
\end{figure}

Compact semiconductor light emitters and detectors operating within the infrared spectrum gained interest within last years as terahertz devices provide the high frequencies needed, e.g., for next generation wireless communications systems; see Refs.\ \cite{nagatsuma_advances_2016, khalatpour2021high, sarieddeen_next_2020, miles_terahertz_2007, tonouchi_cutting-edge_2007}.
Widely tunable intraband (intersubband) transitions in quantum dot (QD) heterostructures can play a crucial role in the development of suitable infrared devices, whereas QD semiconductor devices also show high temperature stability and narrow linewidths; cf.\ \cite{bimberg1999quantum, kirstaedter1994low, arakawa1982multidimensional, grundmann1995ultranarrow}.
In addition, since infrared intraband transitions in quantum well devices are forbidden for light polarized perpendicular to the growth direction (normal incident light), QD-based systems are better suited for infrared sensors (IR photo detectors) \cite{tang2019dual, murata2020infrared, kim2004high, pan1998normal, nagashima2010photodetection, campbell2007quantum}, low-threshold terahertz vertical-cavity surface-emitting lasers \cite{mittelstadt2021terahertz, dmitriev_quantum_2005}, and photovoltaic \cite{beard2010comparing, sablon2011strong, barve2012photovoltaic, sugaya_ultra-high_2011}.\\
In contrast to the polarization of interband transitions in single and stacked QDs (see, e.g., Refs.\ \cite{greif_tuning_2018, yu1999optical, kita2002polarization, kita2003polarization, kita2006artificial, ridha2008polarization, ikeuchi2011multidirectional, yasuoka2011polarization, usman_experimental_2011, usman2012atomistic}), a comprehensive study of the polarization and emission properties of intraband transitions of QD stacks is still missing.
Studies on single QDs and stacks of two QDs (forming a QD molecule) show that electronic intraband transitions in QDs exhibit a fundamentally different spatial emission as interband transitions \cite{jiang1998self, stier_electronic_1999, sheng2001enhanced, sheng2008polarization, adawi2003strong, carpenter2006intraband}.
% genau auf das bisher gezigte eingehen.
%
As well, Sheng \cite{sheng2008polarization} found that the polarization of intraband transitions can be explained by the symmetry of valence band functions, in other words, by directed interactions of the local atomic orbitals.
It was also found that the mixing of these valence band functions into the conduction band states is different for a QD molecule compared to a single QD, giving rise to other transition matrix elements \cite{sheng2001enhanced}.
Based on these preliminary findings on transition matrix elements, we show a comprehensive study of the radiation properties of stacked QDs, guiding the design and growth of such QD structures, potentially making infrared devices more efficient.
Although polarization-resolved transition matrix elements can be derived analogous to interband transitions \cite{he2007growth, saito_optical_2008, ridha2009polarization, usman2011plane, usman_experimental_2011, usman2012atomistic, yuan2018uniaxial}, which in turn can be used to calculate angle-resolved radiation intensities \cite{greif_tuning_2018}, studies investigating the emission directionality of intraband transitions of QD stacks are still lacking.
In this article, we focus on the vertical aspect ratio ($AR$) of the emitter as we found that the emitter's confinement region's aspect ratio has a decisive impact on the emission direction of the intraband transitions similar to that of the interband transitions.
By stacking QDs, ensembles of 30 and more vertically coupled QDs can be formed, changing the emitter's $AR$ and where the interdot coupling strength is adjusted via the separating barrier width \cite{kita2002polarization, sugaya_multi-stacked_2011, li2008shape}.\\
This article is organized as follows: 
First, we outline the method of calculating the QDs electronic structure and the determination of the radiation pattern of CB intraband transitions.
Next, we examine three different series of QDs and highlight the effect of changing the QD stack's morphology on its radiation pattern.
Series A consists of artificial cubodial QDs, which help to identify the impact of the QDs $AR$ on 
emission properties of stacked QDs; see Fig.\ \ref{fig:QD_model}(a).
Series B and C are used to compare our findings with series A with more realistic QD structures:
Here, stacks of InGaAs/GaAs QDs are investigated, mapping the most basic parameters changing the $AR$ of the spatial confinement and the QDs coupling strength, showing that the findings from Series A can be transferred to realistic structures.
In Series B, we demonstrate the impact of the number of QDs stacked. 
Here, for closely stacked QDs, the stack number is varied from a single QD to a stack of ten QDs. 
For series C, for a stack of five QDs, the interdot distance and thus the QD coupling strength is changed from closely stacked into a weak coupling regime until the QDs decouple.
For series B and C, In$_{x}$Ga$_{1-x}$As QDs are modeled as truncated pyramids with a side-wall inclination of $40^{\circ}$, embedded in a GaAs matrix, in agreement with investigations in Refs.\ \cite{blank_quantification_2009, litvinov_influence_2008}.
The model QDs have a base diameter of $17.5\,$nm and a height of $3.4\,$nm, in agreement with experimental investigations of stacked InAs QDs \cite{Inoue2010, ikeuchi2011multidirectional, Kita2011, sugaya_multi-stacked_2011}.
The indium content decreases linearly (see Fig.\ \ref{fig:QD_model}(b)) in agreement with experimental investigations and findings that uniform compositions are not able to reproduce the correct electron-hole alignment and polarization properties for interband transitions \cite{Bruls2002, lemaitre_composition_2004, Fry2000, Barker2000, sabathil2003theory, usman2012polarization}.
\section{Method of calculation}\label{sc:method}
Calculations of the emission directionality of intraband CB transitions in stacks of up to ten QDs are build on an established 8-band $\dotkp$ model; see Refs.\ \cite{grundmann_inas_gaas_1995, stier_electronic_1999, schliwa_impact_2007}.
By using the $\dotkp$ model, electronic structure calculations of extended QD systems are more efficient compared to atomistic models.
Furthermore, long-range interactions like strain and strain-induced piezoelectricity are considered, which usually have to be omitted by atomistic theories; cf.\ Refs.\ \cite{bester_importance_2006-2, bester_effects_2006,stier_modeling_1997, stier_electronic_2000} and references therein.
In the effective mass approximation, single-particle states are represented as a product basis
\begin{align}
\label{eq:expansion}
\ket{\psi} = \sum_{k=1}^{8} \ket{\phi_{k}} \ket{u_{k}} \,,
\end{align}
with the envelope $\ket{\phi_{k}}$ and the atomic-like \note{bandedge ($\Gamma$-point) Bloch functions}
\begin{align}
\label{eq:bloch}
\ket{u} = [\ket{s\uparrow},\ket{x\uparrow},\ket{y\uparrow},\ket{z\uparrow},\ket{s\downarrow},\ket{x\downarrow},\ket{y\downarrow},\ket{z\downarrow}]^{T}    \, ,
\end{align}
where the arrows denote the spin and the $\ket{s}$, $\ket{x},\ket{y},\ket{z}$ denote the $s$- and $p$-type bandedge Bloch functions, respectively.
It is also possible to transform the Bloch basis into another set of eigenfunctions of the Hamiltonian, which represent the states in a conduction (CB) and valence band (VB) Bloch basis; see Refs.\ \cite{chuang1991efficient, enders_kptheory_1995, chuang_physics_2009}. 
For example, considering quantization along the $z$-axis,
\begin{align}
\label{eq:expansion2}
\left[\begin{array}{c} 
\ket{cb_1} \\ \ket{hh_1} \\ \ket{lh_1} \\ \ket{so_1} \\ \vdots 
\end{array}\right] 
= 
\left[\begin{array}{c} 
i\ket{s\downarrow} \\ 
\frac{\ket{x} + i\ket{y}}{\sqrt{2}} \ket{\uparrow} \\ 
\frac{\ket{x} - i\ket{y}}{\sqrt{6}} \ket{\uparrow} + \frac{2\ket{z}}{\sqrt{6}} \ket{\downarrow} \\ 
\frac{\ket{x} - i\ket{y}}{\sqrt{3}} \ket{\uparrow} - \frac{\ket{z}}{\sqrt{3}} \ket{\downarrow} \\ \vdots 
\end{array}\right] \, ,
\end{align}
where the dots represent the functions $\ket{cb_2}$, $\ket{hh_2}$, $\ket{lh_2}$, and $\ket{so_2}$ build from bandedge Bloch functions with opposite spin in Eq.\ \ref{eq:bloch} and $\ket{cb_1} = \ket{u_{cb}\downarrow}$, and so on.
Here, the $\ket{cb}$ denote the CB and the $\ket{hh}$, $\ket{lh}$, and $\ket{so}$  the heavy, light, and splitoff VB states, respectively.
In general, any state $\ket{\psi}$ can be represented in this basis set by projection onto the respective basis functions 
\begin{align}
| \braket{u_k|\psi} |^{2} = | \ket{\phi_{k}} |^{2} \,.
\end{align}
The lattice mismatch of InAs and GaAs (about 7\%) results in a highly strained heterostructure shifting the CB via the deformation potentials $a_c$, $a_v$, and $b$; cf. Refs. \cite{chuang_physics_2009, vurgaftman_band_2001}.
The simplified equation in Ref.\ \cite{chuang_physics_2009} illustrates the strain-induced band coupling for the CB
\begin{align}
%$
V_{\textrm{CB}} = E_{\textrm{CB}} + a_{c}e_{h} 
%$
\end{align}
and the VB
$
V_{\textrm{VB}} = E_{\textrm{VB}} + a_{v}e_{h} \pm \frac{b}{2} e_{b},
$
%\end{align}
where $E_{\textrm{CB}}$ and $E_{\textrm{VB}}$ are the unstrained CB edges, respectively. The $e_{h}$ and $e_{b}$ are the hydrostatic and biaxial strain components, respectively.
Thus, the hydrostatic strain is shifting the CB's energetic position, while
in contrast, the VB is also influenced by the biaxial strain resulting in a splitting of the light- and heavy-hole band.
Consequently, the confinement region and the strain determine the character of the hole ground state, i.e., the leading $\ket{hh}$ and $\ket{lh}$ contributions, respectively, which have a decisive influence on the polarization of the electron-hole decay \cite{greif_tuning_2018}.
In contrast to the hole ground state, the lowest CB state is less affected by the strain (see Ref.\ \cite{enders_kptheory_1995}) and the envelope's symmetry is not changed.
The leading contributions of the CB states is, at least for the InGaAs/GaAs QD systems considered, always $s$-like, but the VB states mix into the states and influence the polarization of the CB intraband transitions and thus their radiation characteristics; cf.\ Refs.\ \cite{chuang_physics_2009, sheng2001enhanced, sheng2008polarization}.
Similar to the change of the VB groundtate's composition considered for interband transitions (see Ref.\ \cite{greif_tuning_2018}), the $\ket{hh}$ and $\ket{lh}$ contribution of the CB groundstate change for stacks of QDs with different heights and spacings.
This is particularly relevant as enlarged $\ket{hh}$ or $\ket{lh}$ contributions result in enhanced horizontal or vertical plus horizontal polarization of the intraband transitions, respectively, owing to the different components from the bandedge Bloch functions; cf.\ Eq.\ \ref{eq:expansion2}.\\
In Fermi's golden rule, the matrix element for a transition between state $\ket{i}$ and state $\ket{j}$ using the expansion in Eq.\ \ref{eq:expansion} is
\begin{multline}\label{eq:matrixelements}
\braket{j|\bm{\epsilon} \cdot \bm{p}|i} = \\
\underbrace{
\sum_{kl=1} 
\delta_{kl} \braket{\phi_{jk}|\bm{\epsilon} \cdot \bm{p}|\phi_{il}}}_{(a)} 
+
\underbrace{
\sum_{kl=1} 
\braket{\phi_{jk}|\phi_{il}} \braket{u_{jk}|\bm{\epsilon} \cdot \bm{p}|u_{il}}}_{(b)} \, ,
\end{multline}
where $\ket{j} = \ket{\psi_{j}}$.
While interband transitions are governed by the second term (b) and are calculated via the optical matrix elements of the bandedge Bloch functions, considering intraband transitions non of both terms might be neglected, whereby the matrix element (a) is calculated from the envelope functions involved, i.e., $\braket{u_{jk}|u_{il}} = 1$.
The emission rate is
\begin{align}\label{eq:rate}
\Gamma_{i,j}(\bm{\epsilon}) \propto 
\Big| [ (a) + (b) ] \Big|^2  \, .
\end{align}
We derive the rate of emitted photons $\Gamma_{i,j}$ as a function of their polarization $\bm{\epsilon}(\phi_{\epsilon},\theta_{\epsilon})$ and their propagation direction $\bm{k}(\phi_{k},\theta_{k})$ by assuming a transversal propagation.
Using a frame of reference $\rho, \varphi, \vartheta$ (in spherical coordinates), with the $z'$-axis aligned to the propagation direction $\bm{k}$, the integrated transition rate for all possible polarizations can be written as a line integral in the $\vartheta= \pi/2$-plane
\begin{align}
\Gamma_{i,j}(\bm{k}) = \sum_{o} \Big[ \int_{0}^{2\pi} d\varphi \Gamma_{i,j}(\varphi, \pi/2) \Big] \, .
\label{eq:Gamma1}
\end{align}
Here, $o$ is the sum over all degenerate intraband states $\ket{\psi}$ considered for the transition.
Throughout this article, we use a spatially resolved emission intensity parametrized in the QD's frame of reference using the condition $\bm{\epsilon} \cdot \bm{k} =0$.
Equation \ref{eq:Gamma1} simplifies to
\begin{align}
\Gamma_{i,j}(\bm{k}) = \sum_{o} \Big[ \int_{\bm{\epsilon} \perp \bm{k}} 
d\varOmega_{\epsilon}^{\textrm{QD}} \Gamma_{i,j}(\bm{\epsilon}) \Big] \, ,
\label{eq:Gamma2}
\end{align}
where $\phi_{k}$ and $\theta_{k}$ define the propagation direction.
If the rate $\Gamma_{i,j}(\bm{k})$ is calculated, each transition is counted several times, thus, the rate is normalized to the total rate of emitted photons.
We calculate the degree of radiation anisotropy (DORA) for the three perpendicular crystal planes.
Here, we use, analogously to the degree of polarization defined in Refs.\ \cite{yu1999optical, kita2002polarization, usman2011plane, usman_experimental_2011}, the $(1\bar{1}0)$- and $(110)$-plane as an indicator for the degree of the QD systems top- and in-plane-emission:
\begin{align}
\textrm{DORA}_{(001)} &=\frac{\Gamma(\bm{k}=[110]) - \Gamma(\bm{k}=[1\bar{1}0])}
{\Gamma(\bm{k}=[110]) + \Gamma(\bm{k}=[1\bar{1}0])}\\
\textrm{DORA}_{(1\bar{1}0)} &=\frac{\Gamma(\bm{k}=[110]) - \Gamma(\bm{k}=[001])}
{\Gamma(\bm{k}=[110]) + \Gamma(\bm{k}=[001])} \label{eq:DORA1-10}\\
\textrm{DORA}_{(110)} &=\frac{\Gamma(\bm{k}=[1\bar{1}0]) - \Gamma(\bm{k}=[001])}
{\Gamma(\bm{k}=[1\bar{1}0]) + \Gamma(\bm{k}=[001])}
\end{align}
%
%
%
% %%%%%%%%%%%%%%%%%%%%%%%%%%%%%%%%%%%%%%%%%%%%%%%%%%%%%%%%%%5
\section{Discussion and Results}%
\label{sc:Discussion}%
As discussed above, interband transitions are governed by the central-cell term Eq.\ \ref{eq:matrixelements}(b).
If we now assume that intraband transitions to be governed by Eq.\ \ref{eq:matrixelements}(a), the trivial conclusion would be that CB $p_{x,y}$-to-$s$-type, $|\ket{\psi_s}|^2$ and $|\ket{\psi_{p_{x,y}}}|^2$, respectively, transitions absorb or emit light polarized within the (001)-plane.
However, previous numerical investigations using eight-band $\dotkp$ models and calculating intraband dipole matrix elements reveal that for QDs, the term Eq.\ \ref{eq:matrixelements}(b) is much larger (about an order of magnitude) than the term (a) calculated from the envelopes involved; cf.\ Refs.\ \cite{jiang1998self, sheng2001enhanced}.
As a consequence, the envelope part is neglected by previous investigations of single QDs and QD molecules (see Refs.\ \cite{sheng2001enhanced, sheng2008polarization}), where an approximation for the dipole moment, which contains the central-cell term solely, is used; see, e.g., Ref.\ \cite{stier_electronic_1999}.
In agreement with previous investigations, we find for the QD series investigated that the transition dipole moment calculated via term Eq.\ \ref{eq:matrixelements}(b) is larger (by about a factor of three to four) than the dipole moment calculated from (a); see Fig.\ \ref{fig:AvsB}.
As such our calculations demonstrate that term (a) has a not negligible impact on the polarization of the CB intraband transitions; therefore, the whole dipole moment is considered within our simulations of the radiation anisotropy via Eq.\ \ref{eq:Gamma2}.
\begin{figure}[tb]
 	\centering
    \includegraphics[]{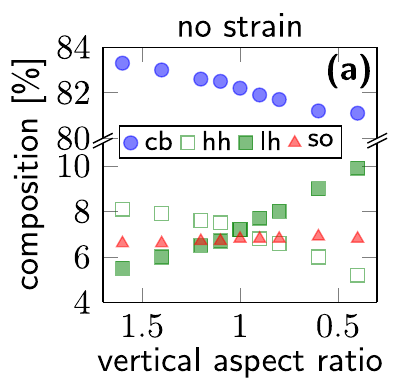}
    \includegraphics[]{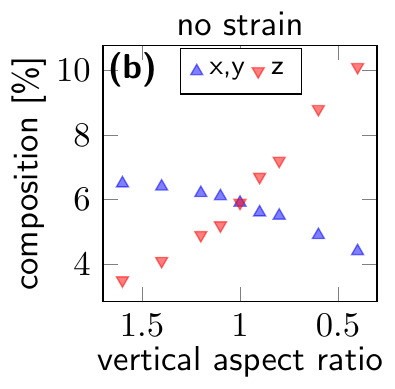}

	\caption{(a) Decomposition of the lowest conduction band state of the cubodial QD series A into the $\ket{cb}$ (blue circles), $\ket{hh}$ (green empty square), $\ket{lh}$ (green filled square), and $\ket{so}$ (red pyramid) basis functions versus vertical aspect ratio.
	(b) Corresponding decomposition into the atomic-like basis functions $\ket{x}$, $\ket{y}$ (blue pyramid), and $\ket{z}$ (red inverted pyramid).
	Both diagrams are calculated excluding effects of strain.
	}
	\label{fig:composition}%
\end{figure}%
\begin{figure*}[tb]
 	\centering
    \includegraphics[]{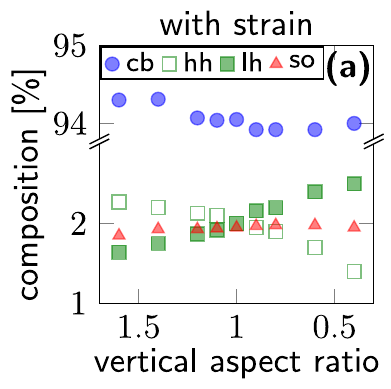}
    \includegraphics[]{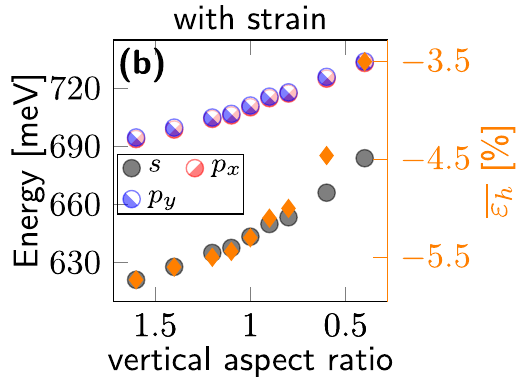}
    \includegraphics[]{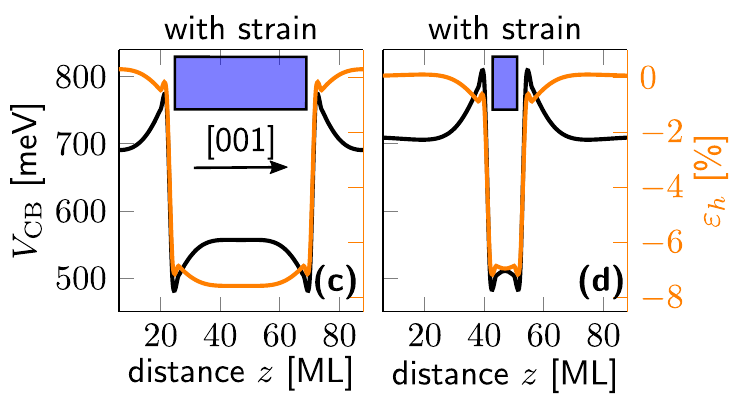}

	\caption{Diagrams are calculated including effects of strain and strain-induced internal fields. 
	(a) Decomposition of the lowest conduction band (CB) state of the cubodial QD series A into the $\ket{cb}$ (blue circles), $\ket{hh}$ (green empty square), $\ket{lh}$ (green filled square), and $\ket{so}$ (red pyramid) basis functions versus vertical aspect ratio ($AR$).
	(b) A plot of the energies for the three corresponding Kramers-degenerate electron states, i.e., the $p_{x,y}$- (semi-filled circles) and $s$-type orbitals (filled circles), where $s$-type state refers to $|\ket{\psi_s}|^2$, likewise for the excited states.
	The right $y$-axis shows the hydrostatic strain expectation value for the lowest CB state (orange diamonds), i.e., 
	$\overline{\varepsilon_{h}} = \braket{\psi_1| \varepsilon_{h} |\psi_1}$.
	(c and d) Plots of the conduction band edge $V_{\textrm{CB}}$ (black) and the hydrostatic strain $\varepsilon_{h}$ (orange) for the cubodial QDs with $AR=1.6$ (c) and with $AR=0.4$ (d).
	}
	\label{fig:comp_with_strain}%
\end{figure*}%
\subsection{The role of aspect ratio.}
With the help of the series of artificial cubic QDs, we show that the geometry of the confinement region alone changes the radiation pattern. 
Therefore, we calculated series A without and with strain.
Figure \ref{fig:composition}(a) shows the different contributions of the VB functions $\ket{hh}$, $\ket{lh}$, and $\ket{so}$ to the lowest CB state without strain.
Changing the QDs $AR$ results in altering contributions of $\ket{hh}$ and $\ket{lh}$ for $AR>1$ and $AR<1$ , while the $\ket{so}$ contribution remains more or less constant.
In contrast to interband transitions, the atomic-like basis functions $\ket{x}$, $\ket{y}$, and $\ket{z}$ in Fig.\ \ref{fig:composition}(b) do not reflect the prevalent confinement region, \note{but as expected,} the $\ket{x,y}$ and $\ket{z}$ fractions are degenerate at $AR=1$.
The contribution of the $\ket{z}$ basis function decreases with growing $AR$, while the $\ket{x,y}$ contribution slightly decreases.
In this regard, intraband transitions show a different characteristic compared to interband transitions.\\
Including strain and strain induced fields, the findings are very similar; see Fig.\ \ref{fig:comp_with_strain}(a).
Here, also an enhancement of the $\ket{hh}$ and a reduction of the $\ket{lh}$ contribution with increasing $AR$ is observed.
As discussed in Sec.\ \ref{sc:method}, the increasing $\ket{hh}$-contribution results in an enhanced in-plane polarization, whereas, on the other hand, a larger $\ket{lh}$-contribution also raises the polarization within the growth direction $z$.
Within the region of confinement, the CB edge shifts to higher energies the larger the hydrostatic strain, and again, the hydrostatic strain becomes stronger the larger the region of confinement; see Figs.\ \ref{fig:comp_with_strain}(c,d) \note{as well as Refs.\ \cite{greif_tuning_2018, mittelstadt2021efficient} for QD stacks}.
As a result, the contribution of the VB functions to the lowest CB state is reduced and, likewise, its leading contribution, $\ket{cb}$-like, is increased. Similarly, an increasing $AR$ reduces the conduction-band-valence-band mixing; see Fig.\ \ref{fig:comp_with_strain}(a).
The shift of the CB edge and the hydrostatic strain as a function of aspect ratio is also reflected in the evolution of the energy eigenvalues of the first three Kramers-degenerate CB states; see Fig.\ \ref{fig:comp_with_strain}(b). The $p$-to-$s$ transition energies shift blue with increasing $AR$, raising the transition matrix element Eq.\ \ref{eq:matrixelements}.
The hydrostatic strain expectation value for the lowest CB state $\overline{\varepsilon_{h}} = \braket{\psi_1| \varepsilon_{h} |\psi_1}$ is increasing with $AR$ reflecting the decreasing energy eigenvalues of the CB states.\\
\begin{figure}[tb]
 	\centering
    \includegraphics[]{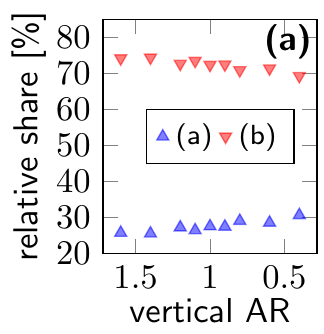}
    \includegraphics[]{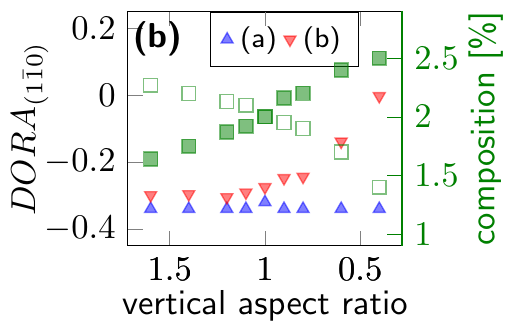}

	\caption{
	(a) Relative share of the term (a) (blue pyramid) and term (b) (red inverted pyramid) contributing to the transition dipole moment in Eq.\ \ref{eq:rate} for the cubodial QD series A versus aspect ratio.
	(b) $\textrm{DORA}_{(1\bar{1}0)}$ of the term (a) (blue pyramid) and (b) (red inverted pyramid) in Eq.\ \ref{eq:rate}, respectively.
	The green axis depicts the contribution of the $\ket{hh}$ (green empty square) and $\ket{lh}$ (green filled square) basis functions from Fig.\ \ref{fig:comp_with_strain}(a). 
	}
	\label{fig:AvsB}%
\end{figure}%
Figure \ref{fig:AvsB}(a) shows the relative shares of the transition dipole moments in Eq.\ \ref{eq:rate}(a) versus (b) in the total dipole moment for the $p_y$-to-$s$ transitions of QDs in series A.
As the $AR$ increases, term (b) gains slightly over (a), with $~2.9$ times greater than (a) for an $AR=1.6$. 
Figure \ref{fig:AvsB}(b) shows the in-plane radiation $\textrm{DORA}_{(1\bar{1}0)}$ determined via Eq.\ \ref{eq:DORA1-10} for the terms (a) and (b), respectively. 
Also, the change in the $\ket{hh}$ and $\ket{lh}$ components of the lowest CB state from Fig.\ \ref{fig:comp_with_strain}(a) is shown to illustrate the change in term (b) from nearly isotropic radiation for $AR<1.0$ to a radiation pattern perpendicular to $(001)$-plane for $AR>1.0$.
Term (a), in contrast, remains constant at a value of $\textrm{DORA}_{(1\bar{1}0)}=-0.34$, caused, as expected, by a strong anisotropic in-plane dipole moment of the $p$-to-$s$ transitions. 
This changes only minimally for an $AR=1.0$, which is caused by the distinct shape of the involved orbitals for the artificial cubic QD at symmetric spatial confinement.
Term (b) alone thus reflects the change in the QD confinement region, resulting in enhanced radiation perpendicular to the growth plane for QDs with $AR>1.0$.\\
\begin{figure}[tb]
 	\centering
    \includegraphics[]{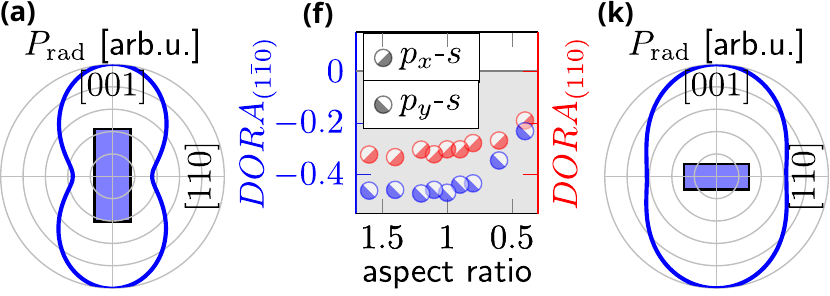}\\
    \includegraphics[]{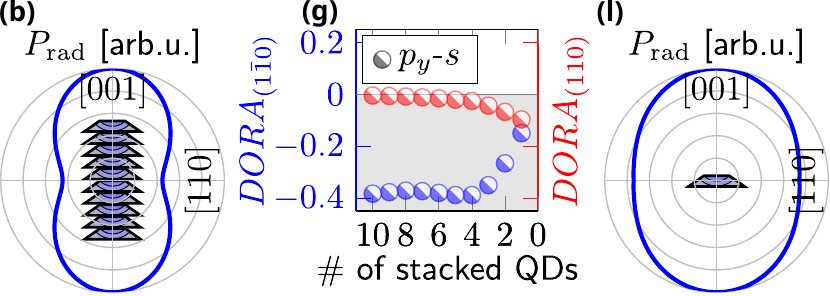}\\
    \includegraphics[]{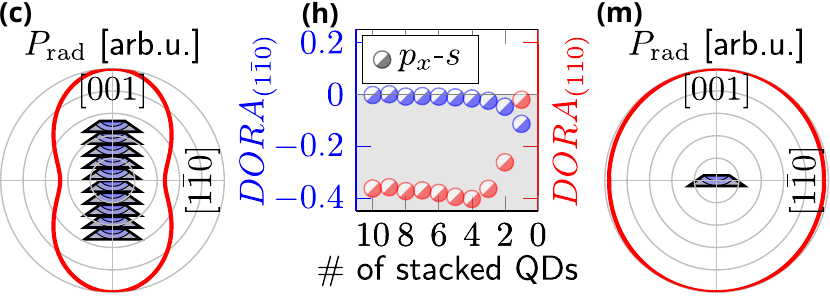}\\
    \includegraphics[]{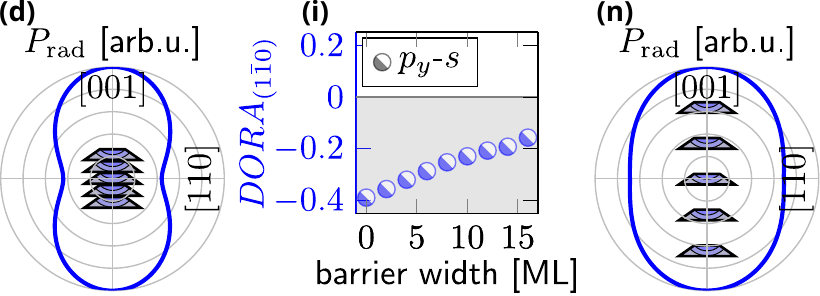}\\
    \includegraphics[]{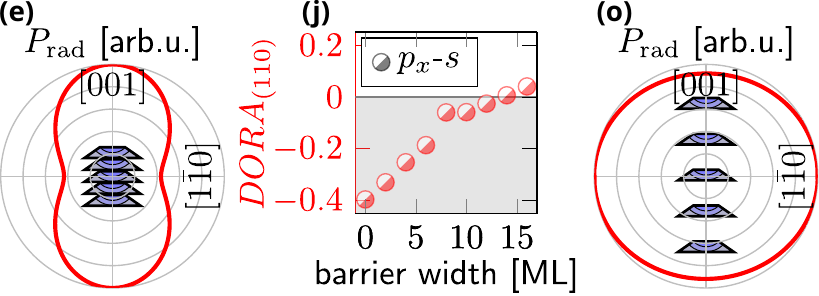}
	\caption{
	(a--e, k--o) normalized radiated power ($P_{\textrm{rad}}$) per solid angle of the first and last QD structures within series A, B, and C, respectively.
	Blue and red radiation patterns correspond to $\textrm{DORA}_{(1\bar{1}0)}$ and $\textrm{DORA}_{(110)}$, respectively.
	(f--j) $\textrm{DORA}_{(1\bar{1}0)}$ (blue semi-filled circle) and $\textrm{DORA}_{(110)}$ (red semi-filled circle) of all QD structures investigated.
	As the symmetry axis of the $p$-type are perpendicular to each other, the blue and red semi-filled circles correspond also to $p_y$- and $p_x$-type CB states, respectively.
	}
	\label{fig:DORA_combined}%
\end{figure}%
The combined impact of terms (a) and (b) on the emission directionality of series A is depicted in Figs.\ \ref{fig:DORA_combined}(a,f,k), which results in a slightly more pronounced DORA of $\textrm{DORA}_{(1\bar{1}0)}\gtrsim -46\%$ for a $AR>1.0$ compared to the individual contributions of terms (a) and (b) shown in Fig.\ \ref{fig:AvsB}(b).
Figure \ref{fig:DORA_combined}(a) shows the radiation pattern within the $(1\bar{1}0)$-plane at an $AR=1.6$, which is highly anisotropic resulting from a strong in-plane polarization for the $p_y$-to-$s$ transition. 
Compared to interband transitions (see Ref.\ \cite{greif_tuning_2018} for $AR<1.0$), intraband transitions at an $AR>1.0$ result in a slightly more pronounced top emission, which makes the contribution of the term (a) in Eq.\ \ref{eq:rate} apparent.
For aspect ratios $AR<1.0$, DORA is increasing, reflecting the growing $\ket{lh}$ contribution to the lowest CB state, but still making top emission more effective than in-plane due to the term Eq.\ \ref{eq:rate}(a); see Fig.\ \ref{fig:DORA_combined}(k).
\begin{figure}[bt]
 	\centering
    \includegraphics[]{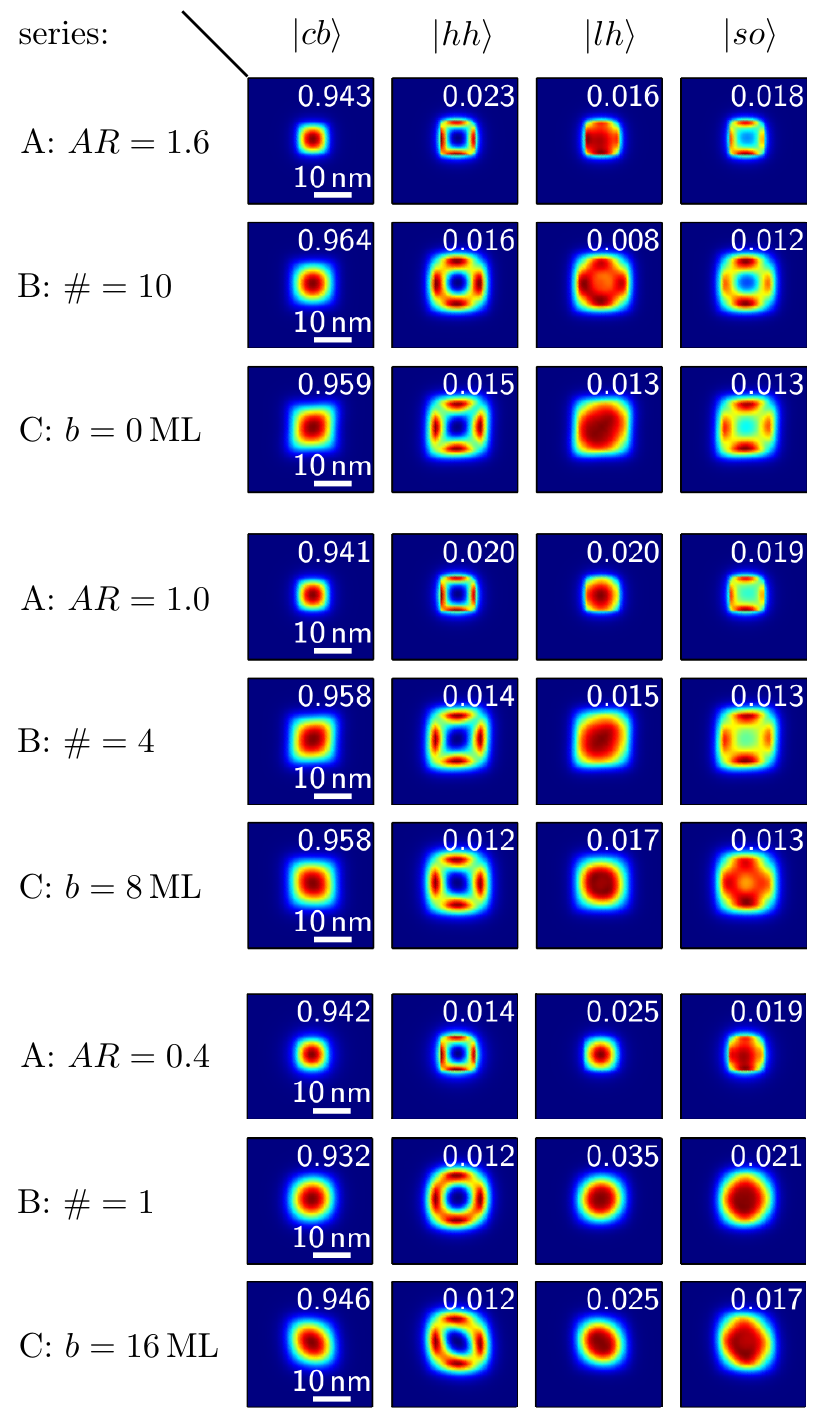}

	\caption{
	The absolute values of the lowest conduction band state basis envelopes $\ket{cb}$, $\ket{hh}$, $\ket{lh}$, and $\ket{so}$ for three selected geometries of series A, B, and C in the $(001)$-plane integrated along the $[001]$-axis and normalized to the maximum in each subfigure. 
	Structures not shown resemble the symmetries of the envelopes depicted, e.g., $AR=1.6$ resembles the symmetries for aspect ratios larger than $1.0$ and $AR=0.4$ for $AR$s smaller than  $1.0$.
	}
	\label{fig:OrbitalsA}%
\end{figure}%
In Figure \ref{fig:OrbitalsA} (series A), the absolute values of the basis functions $\ket{cb}$, $\ket{hh}$, $\ket{lh}$, and $\ket{so}$ orbitals are shown for the vertical aspect ratios $AR=1.6$, $AR=1.0$, and $AR=0.4$ in the $(001)$-plane.
Here, $\ket{cb}$ is the sum of $\ket{cb1}$ and $\ket{cb2}$, likewise for the other three basis functions.
For all $AR$s, the $\ket{hh}$-contribution exhibits a 4-fold symmetry with a central node, resulting in a significant overlap with the CB $p$-type orbitals.
The $\ket{lh}$-contribution changes its symmetry slightly from $s$-like for $AR$s $<1.0$ to a mixture of a 4-fold and $s$-like symmetry for $AR$s $\geq 1.0$ without a central node.
Despite the $\ket{so}$-contribution is barely influenced by changing the $AR$ (cf.\ Figs.\ \ref{fig:composition}(a) and \ref{fig:comp_with_strain}(a)), its symmetry changes from an $s$-like to a 4-fold symmetry with a central node for aspect rations $AR<1.0$ and $\geq 1.0$, respectively.
Due to the change in symmetry of the $\ket{so}$ basis function, the overlap with the $p$-type orbitals increases with aspect ratio, albeit not as pronounced as for the $\ket{hh}$ function.
\subsection{The role of number of stacked QDs.}
In the second series we investigate the emission directionality versus the number of stacked dots.
We choose a barrier width of $b=0$ for this series, as it represents the maximal interdot coupling.
Similarly, by omitting the barrier, the influence of the $AR$ on the emission pattern can be analyzed, also allowing a direct comparison with series A.
Furthermore, experimental findings suggest that columnar QDs may stay in contact with each other, if they were grown with a barrier of a few MLs \cite{kita2002polarization, kita2003polarization,kita2006artificial}.\\
Figures \ref{fig:DORA_combined}(g,h) show the DORA values within the $(1\bar{1}0)$- and $(110)$-plane for the $p_y$-to-$s$ and $p_x$-to-$s$ transitions, respectively.
Beginning with a stack of four QDs, the emission directionality changes with further stacking from nearly isotropic for the single QD with $\textrm{DORA}\approx -2\%$ to a strong top emission with $\textrm{DORA}> -36\%$.
The change from an isotropic to a strongly anisotropic emission direction thus reflects the changing contributions of the VB functions, with the $\ket{lh}$ contribution changing to a predominant $\ket{hh}$ contribution for a stack of four QDs.
We define a collective vertical aspect ratio via the number of stacked QDs, $AR_{\textrm{col}}(n_{\textrm{QD}}) = n_{\textrm{QD}} h_{\textrm{QD}} / (\textrm{average QD width})$, which allows us to compare the $AR_{\textrm{col}}$ with the $AR$ from series A.
As for series A, the emission directionality can be tuned to a strong top emission by stacking QDs to $AR_{\textrm{col}}(n_{\textrm{QD}}) \geq 1.0$; for our series B, the aspect ratio for four QDs is $AR_{\textrm{col}}(4) = 0.92$.
Figures \ref{fig:DORA_combined}(b,c) and (l,m) depict the corresponding emission patterns for a stack of ten QDs and a single QD, respectively, showing a strong top emission for stacks with $n_{\textrm{QD}} \geq 4$.
Figures \ref{fig:DORA_combined}(g,h) also depict the DORA values for radiation parallel to the axis of the $p_{y}$ and $p_{x}$-type orbitals, respectively, showing an isotropic radiation pattern, becoming slightly anisotropic for stacks of QDs smaller than four. 
The symmetries of the VB functions for series B in Fig.\ \ref{fig:OrbitalsA} follow the same pattern as for series A.
\subsection{The role of interdot separation.}
With help of the last series C, we investigate the impact of increasing the interdot barrier $b$ on the emission directionality of stacked QDs.
If QDs are stacked closely with $b=0$\,MLs, a strong top emission with DORA values of $\approx -40\%$ are observed for a stack of five QDs; cf.\ Fig.\ \ref{fig:DORA_combined}(i,j).
Figures\ \ref{fig:DORA_combined}(d,e) and (n,o) depict the radiation pattern for $b=0$\,MLs and $b=16$\,MLs, respectively.\\
By increasing the interdot separation, the $\ket{hh}$- and $\ket{lh}$ contributions shift towards the single QD case of Series B, or $AR=0.4$ for series A.
As shown in Ref.\ \cite{greif_tuning_2018}, the band structure and strain distribution is almost identical to an uncoupled single QD.
In contrast to series B, the emission directionality changes linearly from a strong top to a more isotropic emission of DORA $\approx -15\%$ and $\approx 4\%$, respectively; cf.\ Fig.\ \ref{fig:DORA_combined}(i,j).
The envelopes in Fig.\ \ref{fig:OrbitalsA} show identical symmetries as for series B, despite the $\ket{so}$-contribution at $b=8$\,MLs showing a state changing from an $s$-like to a 4-fold symmetry.
We note, however, that an anti-binding $s$-type state forms as the QDs separation is increased, which would make measuring the radiation characteristics of the $p$-$s$ transitions even more difficult for stacks with large barriers; cf.\ Refs.\ \cite{mittelstadt2021efficient}.
\section{Conclusion}
We demonstrate that the vertical emission of intraband transitions of stacked QDs can be tuned via the structure's vertical aspect ratio.
With the help of an artificial cuboidal QD, we show that the minor contributions from the central-cell part within the momentum matrix element change with aspect ratio enhancing or weakening the emission anisotropy.
In contrast to interband transitions, the strain has a subordinate role and the change of the aspect ratio itself results in a more dominant $\ket{lh}$/$\ket{hh}$ contribution of the lowest CB state.
In addition, we find that the matrix elements among the envelope functions have a significant contribution to the emission direction of stacked QDs and cannot be neglected.\\
We investigate two series of QDs, investigating the impact of (i) the number of stacked QDs on the radiation pattern and (ii) the coupling strength between stacked QDs.
Here we show that the results for the artificial QD can be transferred to realistic QD structures.
In QD stacks, stacking QDs leads to a larger aspect ratio, thus enhancing the top emission for the CB intraband transition, although this is only beneficial up to the isotropic with an $AR$ of 1.0, as further stacking then provides no additional enhancement.
Moreover, we show that the vertical emission is enhanced with increasing AR, vice versa to the interband transitions in QDs.
By increasing the interdot distance in QD stacks, i.e., weakening the QD coupling strength, the emission pattern converges to a single QD, reducing the vertical emission to a more isotropic pattern.\\
Similar to interband transitions, our calculations show qualitative changes in the radiation pattern of QD stacks and provide qualitative values of the degree of radiation anisotropy (DORA) for the parameters considered. 
In addition, we provide insight into the contribution of the envelope functions versus the central cell fraction to the emission rate and pattern.
Our calculations also consider the contribution of the envelope functions to the emission pattern, which has been neglected in previous studies of the polarization of intraband transitions as its contribution is comparatively small.
Therefore, our studies provide new insight into the radiation anisotropy of QD intraband transitions, especially for stacked QDs, which may be beneficial for designing future infrared devices with enhanced in-plane or top emission polarization.
\begin{acknowledgments}
The authors would like to thank Dirk Ziemann for reading the manuscript and for the fruitful and constructive discussions.
\end{acknowledgments}
\bibliography{lib_korr}% Produces the bibliography via BibTeX.

\end{document}